\documentclass[superscriptaddress,twocolumn,secnumarabic,
 amssymb,amsmath,nobibnotes,aps,prd,showkeys,showpacs]{revtex4}

\usepackage{graphicx}
\usepackage{bm}%
\newcommand{\be}{\begin{equation}}
\newcommand{\ee}{\end{equation}}
\newcommand{\bea}{\begin{eqnarray}}

\newcommand{\eea}{\end{eqnarray}}

\begin{document}
%opening
\title{How to observe a non-Kerr spacetime}

\author{Theocharis A. Apostolatos}
\affiliation{Section of Astrophysics, Astronomy, and Mechanics,
Department of Physics, University of Athens, Panepistimiopolis Zografos GR15783,
Athens, Greece}

\author{Georgios Lukes-Gerakopoulos}
\affiliation{Academy of Athens, Research Center for Astronomy,
Soranou Efesiou 4, GR-11527, Athens, GREECE}
\affiliation{Section of Astrophysics, Astronomy, and Mechanics,
Department of Physics, University of Athens, Panepistimiopolis Zografos GR15783,
Athens, Greece}

\author{George Contopoulos}
\affiliation{Academy of Athens, Research Center for Astronomy,
 Soranou Efesiou 4, GR-11527, Athens, GREECE}

\begin{abstract}
We present a generic criterion which can be used in
gravitational-wave data analysis to distinguish an
extreme-mass-ratio inspiral into a Kerr background spacetime from
one into a non-Kerr background spacetime. The criterion exploits the
fact that when an integrable system, such as the system that
describes geodesic orbits in a Kerr spacetime, is perturbed, the
tori in phase space which initially corresponded to resonances
disintegrate so as to form the so called Birkhoff chains on a
surface of section, according to the Poincar\'{e}-Birkhoff theorem.
The KAM curves of these islands in such a chain share the same ratio
of frequencies, even though the frequencies themselves vary from one
KAM curve to another inside an island.
%C However
On the other hand, the KAM curves, which do not lie in a Birkhoff chain, do not
share this characteristic property.
%C according to KAM theorem.
Such a temporal constancy of the ratio of frequencies during the evolution of
the gravitational-wave signal will signal a non-Kerr spacetime which could then
be further explored.
\end{abstract}

\maketitle

%%%%%%%%%%%%%%%%%%%%%%%%%%%%%%%%%%%%%%%%%%%%%%%
\section{Introduction}
\label{sec:1}
%%%%%%%%%%%%%%%%%%%%%%%%%%%%%%%%%%%%%%%%%%%%%%%

While ground-based gravitational-wave detectors are already in
operation, and are trying to detect gravitational waves from stellar
mass compact objects, LISA  --the space-borne detector \cite{LISA}
which is planned to be launched during the forthcoming decade-- is
expected to observe much more massive sources of gravitational waves
with high signal-to-noise ratio. Such signals will offer us an
opportunity to map the strong field around massive astrophysical
objects \cite{Ryan95,GlamBaba05}, and to test the conventional
wisdom, partly supported by astrophysical observations
\cite{Menouetal99,Seoaetal07}, according to which the massive
compact objects harbored at galactic centers should be highly
spinning Kerr black holes. Such an astrophysical fact is further
enforced by the no-hair theorem, which states that when a black hole
is formed, its multipole moments, and consequently its gravitational
field, are determined by only two parameters; its mass and spin
(assuming its charge is negligible). In this paper we suggest a
clear and generic observable signal that distinguishes an
extreme-mass-ratio inspiral (EMRI) \cite{EMRI} related to a non Kerr
black hole spacetime from a corresponding Kerr one.

Our suggestion comes from the fact that a Kerr spacetime leads to an
integrable system that describes geodesic orbits, while any other
generic, stationary, and axisymmetric non-Kerr spacetime is not
expected to be integrable. Thus assuming that a slightly perturbed
Kerr metric, that supposedly describes the neighborhood of an
axisymmetric rotating compact object, is governed by a non
integrable system of geodesic equations of motion, any qualitative
new characteristics of the orbit
%A
of the test body, which could in principle be observed
%A
through gravitational waves, can be used to identify such a source.

According to the KAM theorem \cite{Arno} for dynamical systems almost all KAM
tori in the phase space of a perturbed integrable system are not destroyed; they
simply become slightly deformed. However, among the KAM tori of the initial
integrable system, there are the so called resonant tori  that are characterized
by commensurate ratios of frequencies. In the perturbed system these tori
%C get
disintegrate, and according to the Poincar\'{e}-Birkhoff theorem
\cite{BirkPoin} they form a chain of islands, on a surface of
section, inside which the ratio of the corresponding frequencies
remains equal to the rational number of the corresponding initial
resonant torus. The width of this chain of islands is a
monotonically growing function of the perturbative parameter that
%C adjusts
measures the system's deviation from the corresponding integrable
one, at least for small perturbations.

In our case, an EMRI in a perturbed Kerr spacetime will be described by an
adiabatically changing geodesic orbit, which will sweep a finite range of KAM tori
%C on
in the course of time, while the frequencies of the  orbital
oscillations on the polar plane (the plane that passes through the
axis of symmetry of the massive object and rotates along with the
low-mass body) will change continuously. When the system enters a
Birkhoff chain of islands on a surface of section, the ratio of
frequencies will remain constant, while the frequencies will
continue to change. Therefore the appearance of a plateau in the
ratio of frequencies during the evolution of an EMRI will definitely
signal the presence of a non-Kerr spacetime. These frequencies will
be encoded in the gravitational wave that is radiated from the
corresponding source. Thus they can in principle be monitored. The
question is how probable is for an orbit to cross such a chain of
islands during its evolution that is monitored by the LISA detector.
We argue \cite{Aposetal} that this happens quite often, since the
orbits of EMRI's that
%C develops %A
develop in the neighborhood of  the supermassive compact object at
galactic centers are initially quite eccentric and inclined
\cite{Babaetal}.
%A
Depending on the value of its physical parameters, the system will
eventually enter the region of such an island and then the plateau
in the ratio of frequencies will show up. Furthermore, since the
rational ratios of frequencies are dense in any finite interval,
during the whole inspiral phase there is a non-negligible
probability for observing more than one plateaus in frequency
ratios. It should be noted though that only the low-number
resonances are probable to be observed since the width of the
corresponding islands is usually much lower for the higher
resonances.

In order to exhibit our
%A
criterion for observing a non-Kerr spacetime, we have used a
Manko-Novikov (MN) metric \cite{MankNovi,Gairetal} as an example of
a non-Kerr metric. The specific MN metric is characterized by one
parameter $q$, besides the mass and the spin parameters, which
measures the deviation of its quadrupole moment from the
corresponding Kerr metric. This particular metric was analyzed in
\cite{Gairetal} and was found to behave like an integrable system
with respect to regularity of its curves in the outer allowed
region of phase space. Based on the general applications of the KAM
theorem, we performed a more thorough analysis of the orbits in the
same metric and found that the expected Birkhoff chains of islands
are present on a surface of section, although they are very thin to
be observed in a coarse study of orbits.

Also we investigated the effect on the ratio of frequencies of polar
oscillations when an orbit is moving adiabatically in and out of
such an island while the corresponding  EMRI radiates energy and
angular momentum away. We have shown that the plateau in the ratios
of frequencies could last for a few hours to a few weeks, depending
on the masses involved in the EMRI, as well as the value of the $q$
parameter of the background spacetime. Actually, the aforementioned
plateaus will be more distinctly observed in lower mass-ratios
($\mu/M \lesssim 10^{-5}$ where $\mu$, $M$ are the reduced and total
mass of the binary) since then the system spends more time within an
island. These effective plateaus could be somehow integrated in the
schemes used in data analysis of LISA to look for non-Kerr EMRI's.

Finally, we should emphasize that the existence of a plateau should
be a generic result for any perturbed Kerr metric, therefore the MN
metric
%A
which was assumed for the background metric is an example that does
not imply any restrictions in the general application of the
proposed observable criterion.

%%%%%%%%%%%%%%%%%%%%%%%%%%%%%%%%%%%%%%%%%%%%%%%
\section{Resonances in a non-Kerr MN spacetime}
\label{sec:2}
%%%%%%%%%%%%%%%%%%%%%%%%%%%%%%%%%%%%%%%%%%%%%%%

The ``bumpy'' spacetime, which was used by in \cite{Gairetal} to
explore the orbits in a non-Kerr metric, is a family of solutions of
the Einstein equations in vacuum, that has been built on the
foundations of a Kerr metric and thus it can be transformed into a
pure Kerr metric \cite{MankNovi}. For the specific MN metric this
could actually be done by setting the $q$ parameter (the quadrupole
deviation parameter) equal to zero. The higher mass and mass-current
multipoles are also affected by $q$ \cite{Gairetal}.

The Kerr metric is a very special metric for two main reasons. (a)
It is a highly symmetric metric since apart from the three integrals
of geodesic motion that all axially symmetric, stationary, and
asymptotically flat metrics share, it is characterized by an extra
integral of motion, the so called Carter constant. (b) It is a
metric related to realistic physical objects. According to the ``no
hair theorem'', all spinning objects that collapse to form a black
hole are described by such a metric. Since highly compact objects
that move very close to each other are very powerful sources of
gravitational radiation, an EMRI of a stellar-mass compact object
around a massive Kerr black hole is a highly promising source for
the LISA detector.

%%%%%%%%%%%%%%%%%%%%%%%%%%%%%%%%%%%%%%%%%%%%%%%%%%%%%%%%%%%%%%
\begin{figure}[h]
%PUT FIG2 OF PRD
 %\centerline{\includegraphics[width=16pc] {FigIntReg.eps}}
 \centerline{\includegraphics[width=14pc] {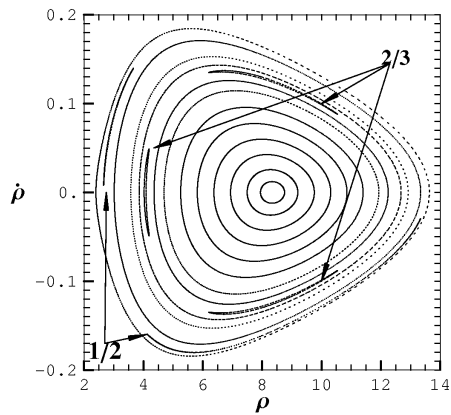}}
 \caption{ The surface of section of the outer region of allowed orbits on the
 $\rho,~\dot{\rho}$ plane for the parameter set $E=0.95,~L_z=3 M,~\chi=0.9,
  ~q=0.95$ of MN
 (for the definition of these quantities see \cite{Gairetal}).}
 \label{fig:1}
\end{figure}
%%%%%%%%%%%%%%%%%%%%%%%%%%%%%%%%%%%%%%%%%%%%%%%%%%%%%%%%%%%%%%

%A
In order to look for qualitative new characteristics in
%C other
non-Kerr compact sources that could exist in nature, we have
computed numerically the
%A
geodesic orbits in a MN metric. Since this
is a metric which is built on the basis of a Kerr metric, while it
does not share its special
%C symmetries
%A
symmetry that leads to Carter constant, it could be considered as a
perturbed Kerr; therefore it is not expected to be described by an
integrable system, in contrast to what happens in a Kerr spacetime.
The fact that this is not an integrable system could already be
deduced from the fact that the orbits in the inner allowed region of
the MN spacetime exhibit
%C ergodicity
chaotic behavior, at least in all cases studied in \cite{Gairetal}.
%A
On the other hand the apparent regularity of orbits in the outer
region is just an outcome of
the
KAM theorem,
%A
as noted in \cite{Gairetal}. In this region (where the gravitational
field is weaker) the MN spacetime looks like a perturbation of Kerr
and thus most of KAM tori just modify their shape. However, the
resonant tori disintegrate and intersect a surface of section
forming regions of finite thickness, instead of single closed KAM
curves. These regions form
%C a chain
sets of islands, known as Birkhoff chains of islands; each one of
them bounded externally and internally by KAM curves \cite{Conto}.
The existence of such chains of islands is characteristic of a
system that is nearly integrable. A thorough analysis of geodesic
orbits in the MN spacetime revealed two such Birkhoff chains of
islands that correspond to
%A
the resonances $2/3$ and $1/2$. Discovering such islands in a
Poincar\'{e} surface of section is not an easy task since these
islands are not wide enough (cf.~Fig.~\ref{fig:1}); a very fine
sweep of initial conditions is needed to get such an island.
Fortunately, there is an alternative tool to approach the
corresponding orbit much faster, the ``rotation number''. This
number
%C could
can be easily computed for every orbit and by changing the initial
conditions and monitoring  the corresponding rotation number, we
could arrive at the desirable rational value of the rotation number,
by a method analogous to a root-finding algorithm through bisection.

A phase orbit in an integrable system (like the Kerr case) is wound around a
non-resonant torus filling the whole torus densely in the course of time, while
for a resonant torus the orbit
% repeats
is periodic, repeating itself after a few windings \cite{Arno}.
Thus, on a surface of section which intersects transversally the
corresponding tori (we have used the plane $z=0$, on which we mark
the intersecting points when $\dot{z}>0$), the crossing points
constitute a  set of points that densely fill a closed curve in the
former case, and a finite number of fixed points in the latter case.
This number is the number of oscillations along the spatial axis
that intersects transversally the surface of section (the $z$-axis
in our case) that correspond to a finite number of oscillations
along the other spatial axis (the $\rho$ axis). When the system
deviates from being integrable, these fixed points of the periodic
orbit evolve to an equal number of Birkhoff islands.

By means of the sequence of crossing points described above it is
easy to define the aforementioned rotation number. Either for an
integrable or for a non integrable system of 2 degrees of freedom,
this is defined as follows: Let ${\cal A}$ be the fixed point on a
surface of section and ${\cal B}_i$ be the $i$-th point of
intersection of the phase orbit with the surface of section. The
vector $\overrightarrow{{\cal A}{\cal B}_i}$ rotates through an
angle $\Delta\theta_{i}$ as it moves from the $i$-th intersecting
point to the next one. The rotation number is $\nu=\displaystyle
\frac{1}{2 \pi} \lim_{N \rightarrow \infty} \frac{1}{N}
\sum_{i=1}^{N} \Delta\theta_{i}$ (for a review see \cite{Conto}). As
explained above the rotation number is the fraction of the two
corresponding fundamental frequencies of the system, each multiplied
by an integer. Thus for a geodesic orbit in a MN metric it will be
%\be
$\nu=\frac{m f_{\rho}}{n f_{z}}$
%\ee
with $m,n$ integer numbers. Whenever the two frequencies are
commensurate to each other, the rotation number is rational.
%A
Actually the resonant tori of the integrable system that
%A leads
form  Birkhoff islands on a surface of section
when the system gets perturbed, are the ones with commensurate
frequencies, and consequently with rational rotating numbers.
%A
Thus we
can alter the initial conditions
%A to
until we obtain a specific rational value
for $\nu$
%A
and thus
%A
reveal the location of a Birkhoff island. For a
given perturbed system, the actual
%A
width of the Birkhoff islands
is often higher for lower-integer ratios of frequencies,
%C \cite{Efth}, %A a comma
while for higher-integer ratios the islands are so thin, that they
are very difficult to be discerned. In our case of geodesic orbits
in MN we have searched for Birkhoff islands when $f_{\rho}/f_{z}=2$,
or $1$, which correspond to $\nu=2/3$, or $1/2$, respectively. On
the other hand the thickness of the particular islands grows with
the perturbation parameter $q$, though not linearly, in agreement
with simple perturbed integrable Hamiltonian toy models that have
already been studied in the literature \cite{Conto,Conto1,Efth}.

%%%%%%%%%%%%%%%%%%%%%%%%%%%%%%%%%%%%%%%%%%%%%%%%%%%%%%%%%
\section{Evolving orbits due to radiation reaction}
\label{sec:3}
%%%%%%%%%%%%%%%%%%%%%%%%%%%%%%%%%%%%%%%%%%%%%%%%%%%%%%%%%

\begin{figure}
 \centerline{\includegraphics[width=14pc] {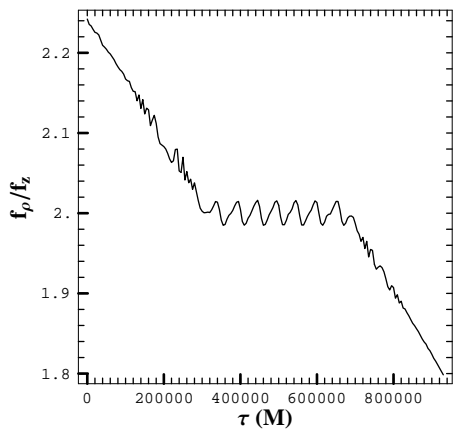}}
%\centerline{\includegraphics[width=16pc] {07.eps}}
 \caption{ The evolution of the ratio of fundamental frequencies
 $f_\rho/f_z$ for a MN EMRI with $q=0.95$ and ratio of masses $\mu/M=5\times10^{-5}$.
 The initial conditions were chosen to get quite a long plateau.
 The fundamental frequencies were computed numerically from the Fourier analysis of
 the orbital motion in time intervals of length $5000 M$. The oscillations
 of the ratio --mainly at the plateau interval-- are artifacts due to
 Fourier analyzing a finite part of the orbit.
 }
 \label{fig:plateau}
\end{figure}

Each chain of Birkhoff islands, separate the surface of section in an interior
and an exterior region. While one moves from the former to the latter region,
the rotation number, and consequently the ratio of frequencies, drops
monotonically, except if we land on the thin window of an island;
%C  itself
then the ratio of frequencies does not change, even though the
frequencies themselves change. Thus by analyzing the spectrum of a
gravitational wave signal and monitoring the ratio of the
fundamental frequencies (cf.~Fig.~\ref{fig:plateau}), whenever a
plateau shows up, it will signal a non-Kerr type of background. The
plateau will be more prominent (i) if the deviation from Kerr is
larger, (ii) if the ratio of mass is more extreme, since then the
deviation from geodesic orbit,
%C
due to energy and angular momentum loss,
is slower and the system needs more time to cross an island,
and (iii) the island corresponds to a resonance of low-integer ratios,
since these resonances correspond to wider islands.

%%%%%%%%%%%%%%%%%%%%%%%%%%%%%%%%%%%%%%%%%%%%%%%%%%%%%%%%%
%%%%%%%%%%%%%%%%%%%%%%%%%%%%%%%%%%%%%%%%%%%%%%%%%%%%%%%%%%

We have run such adiabatic orbits in MN
%C spacetime
spacetimes numerically, following approximate formulae for the
evolution of energy and angular momentum along the axis of symmetry
(for a generic spacetime there is no reliable way to compute the
losses due to radiation) \cite{GairGlam}. Our numerical explorations
show that these plateaus are visible for ratios of masses below some
threshold that depends on the resonance we are dealing with and the
value of $q$ assumed for the MN metric. For example for $q=0.95$ the
higher ratio of masses for which we
%A
can tell the presence of such a plateau is $\mu/M
\cong 10^{-4}$. If the ratio of masses is above these thresholds,
the system evolves so fast that the corresponding
%C plateau is simply %A
plateaus are not discernible. Below the threshold, the actual duration of the
plateau is more extended for lower values of $\mu/M$, although it varies a lot
depending on the specific trajectory of the phase orbit through the Birkhoff
island. The aforementioned thresholds are actually within the relative range of
masses expected for an EMRI signal detectable by LISA \cite{Dras06}.

Another crucial point with respect to observability of the plateau effect is the
following: Since the chains of Birkhoff islands are numerous, and the phase
orbit is forced to move adiabatically from one KAM curve to the next, it is
forced to pass through many such islands. Thus while the effect is always
present, an actual system will exhibit an unambiguous plateau whenever it has
sufficient time to cross
%C an
a strong resonance, like the $2/3$ or the $1/2$ one. This happens if the
geometric characteristics of the orbit when it enters the window of sensitivity
of LISA are such that during the evolution of the orbit, it will cross one of
these resonances. Thus the orbit should have initial eccentricity and
inclination within a suitable range. These ranges are quite wide; therefore it
is anticipated that a large fraction of such suitable EMRI's will leave their
imprints on their signal through an apparent plateau of the ratio of the
observed frequencies. Moreover even if the evolution of the signal is such that
it correlates well with a Kerr EMRI, we could focus our search in the particular
period when the ratio of the corresponding peaks in the spectrum that are
related to the radial and precessional oscillation are close to a resonant ratio
($f_{\rho}/f_z=2$ for the $2/3$ resonance and $1$ for the $1/2$ resonance). A
statistically important persistence of such a ratio would be a clear ``smoking
gun'' for a  non-Kerr metric.

Finally, we should note that a possible positive signal from a
non-Kerr EMRI, could be further explored for other consequences of
such a peculiar source, like the instabilities that are expected to
show up before the corresponding final plunge \cite{Gairetal}, or a
possible transit of the orbit to
%C ergodic
chaotic behavior through entrance in the interior region of allowed
orbits of a MN-like metric \cite{Gairetal, Aposetal}.

%%%%%%%%%%%%%%%%%%%%%%%%%%%%%
\begin{acknowledgments}
T.~A.~acknowledges the research funding program ``Kapodistrias'' of
ELKE (Grant No 70/4/7672). G.~L.-G. was supported in part by the IKY
scholarships and by the Research Committee of the Academy of Athens.
\end{acknowledgments}
%%%%%%%%%%%%%%%%%%%%%%%%%%%%%


\begin{thebibliography}{plain}

\bibitem{LISA}
P.~Bender et al., 'LISA Pre-Phase A Report' 2nd Ed., (1998).
%
\bibitem{Ryan95}
F.~D.~Ryan, Phys.~Rev.~D, \textbf{52}, 5707 (1995).
%
\bibitem{GlamBaba05}
K.~Glampedakis, S.~Babak, Class.~Quant.~Grav., {\bf 23}, 4167 (2006).
%
\bibitem{Menouetal99}
K.~Menou, E.~Quataert, R.~Narayan, 'Recent Developments in Theoretical and
Experimental General Relativity, Gravitation and Relativistic Field Theories'
(World Scientific Publishers, 1999)
%
\bibitem {Seoaetal07}
P.~Amaro-Seoane, J.~R.~Gair, M.~Freitag, M.~C.~Miller, I.~Mandel, C.~J.~Cutler,
S.~Babak, Class.~Quantum Grav., \textbf{24}, R113 (2007).
%
\bibitem{EMRI}
L.~Barack, C.~Cutler,
Phys.~Rev.~D, \textbf{75}, 042003 (2007).
%
\bibitem{Arno}
V.~I.~Arnold, `Mathematical Methods of Classical Mechanics', 2nd Ed., (Springer, 1989).
%
\bibitem{BirkPoin}
H.~Poincar\'{e}, Rend.~Circ.~Mat.~Palermo, \textbf{33}
375 (1912), and G.~D.~Birkhoff, Trans.~Am.~Math.~Soc.,
\textbf{14}, 14 (1913).
%
\bibitem {Aposetal}
T.~A.~Apostolatos, G.~Lukes-Gerakopoulos, J.~Deligiannis,
G.~Contopoulos, work in progress.
%
\bibitem{Babaetal}
S.~Babak, H.~Fang, J.~R.~Gair, K.~Glampedakis, S.~A.~Hughes,
Phys.~Rev.~D, \textbf{75}, 024005 (2007), Erratum-ibid. \textbf{77}, 04990 (2008).
%
\bibitem {MankNovi}
 V.~S.~Manko, I.~D.~Novikov,
 Class.~Quant.~Grav., {\bf 9}, 2477 (1992).
%
\bibitem {Gairetal}
 J.~R.~Gair, C.~Li, I.~Mandel,
 Phys.~Rev.~D, {\bf 77},~024035 (2008).
%
 \bibitem {Conto}
 G.~Contopoulos,
`Order and chaos in dynamical Astronomy',
 (Springer, 2002)
%
 \bibitem {Conto1}
 G.~Contopoulos, Bull.~Astron. (3), \textbf{2}, 223 (1967).
%
\bibitem{Efth}
C.~Efthymiopoulos, Ce.~M.~D.~A., \textbf{102}, (1-3), 49 (2008).
%
\bibitem{GairGlam}
J.~R.~Gair, K.~Glampedakis, Phys.~Rev.~D, \textbf{73}, 064037 (2006).

%\bibitem {Voglis98}
% Voglis N., Efthymiopoulos C.: 1998,
% {\it J. Phys. A: Math. Gen.\/},{\bf 31},~2913
%
\bibitem {Dras06}
S.~Drasco, Class.~Quant.~Grav., \textbf{23}, S769 (2006).



\end{thebibliography}
\end{document}